\documentclass[3p,times]{elsarticle}

\usepackage{ecrc}

\usepackage{epstopdf}

\volume{00}

\firstpage{1}

\journalname{Nuclear Physics A}

\runauth{M.~J.~Tannenbaum et al.}

\jid{nupha}

\jnltitlelogo{Nuclear Physics A}

\usepackage{graphicx}
\usepackage{amsmath,amssymb}

\usepackage{xspace}     

\newcommand{\pt}{\mbox{$p_T$}\xspace}

\newcommand{\Npart}{\mbox{$N_{\rm part}$}\xspace}
\newcommand{\Ncoll}{\mbox{$N_{\rm coll}$}\xspace}
\newcommand{\Nch}{\mbox{$N_{\rm ch}$}\xspace}
\newcommand{\Et}{\mbox{${\rm E}_T$}\xspace}

\newcommand{\sqs}{\mbox{$\sqrt{s}$}\xspace}
\newcommand{\sqsn}{\mbox{$\sqrt{s_{_{NN}}}$}\xspace}

\newcommand{\Nqp}{\mbox{$N_{qp}$}\xspace}

\def\lsim{\raise0.3ex\hbox{$<$\kern-0.75em\raise-1.1ex\hbox{$\sim$}}}
\def\gsim{\raise0.3ex\hbox{$>$\kern-0.75em\raise-1.1ex\hbox{$\sim$}}}
\def\mean#1{\langle#1\rangle}

\def\Journal#1#2#3#4{{#1}{\ #2} (#4) #3 }

\def\NPB{{Nucl. Phys. B}}
\def\PLB{{Phys. Lett. B}}

\def\PL{Phys. Lett.\ }
\def\PRL{Phys. Rev. Lett.\ }
\def\PRD{{Phys. Rev. D}}
\def\PRC{{Phys. Rev. C}}

\def\ZPC{{Z. Phys. C}}
\def\ARNPS{{Annu. Rev. Nucl. Part. Sci.\ }} 
\def\RMP{Rev. Mod. Phys.\ }

\begin{document}

\begin{frontmatter}



\title{Transverse energy (\Et) distributions at mid-rapidity in $p$$+$$p$, $d$$+$Au and Au$+$Au collisions at \sqsn=200 GeV and implications for particle production models }

\author{M.~J.~Tannenbaum  (for the PHENIX\fnref{col1} Collaboration)}
\fntext[col1] {A list of members of the PHENIX Collaboration and acknowledgements can be found at the end of this issue.}
\address{Physics Department, Brookhaven National Laboratory, Upton, NY 11973-5000, USA}




\begin{abstract}
Measurements of the mid-rapidity transverse energy distribution $d\Et/d\eta$ are presented for  $p$$+$$p$, $d$$+$Au, and Au$+$Au collisions at \sqsn=62.4--200 GeV. The \Et distributions are compared with the number of participants, \Npart, the number of constituent-quark participants, \Nqp, and the number of color-strings (Additive Quark Model --- AQM) calculated from a Glauber model. For Au$+$Au, $\mean{ d\Et/d\eta}/(0.5 N_{\rm part})$ increases with $N_{\rm part}$, while $\mean{ d\Et/d\eta}/N_{qp}$ is approximately constant vs. centrality for $\sqsn \geq 62.4$ GeV. This indicates that the two 
component ansatz, $d\Et^{\rm AA}/d\eta=(d\Et^{\rm pp}/d\eta)\ [(1-x)\, \Npart/2 + x\, \Ncoll]$, which has been used to represent \Et distributions, is simply a proxy for \Nqp, and that the \Ncoll term does not represent a hard-scattering component in \Et distributions. The $d\Et/d\eta$ distributions of $d$$+$Au, and Au$+$Au are calculated from the measured $p$$+$$p$ \Et distribution using two models (AQM and \Nqp) that both reproduce the Au$+$Au data. For the asymmetric $d$$+$Au system, the \Nqp model reproduces the data while the AQM does not. 
\end{abstract}


\end{frontmatter}



\section{Introduction}
\label{sec:intro}
Recent PHENIX measurements of mid-rapidity transverse energy distributions   
$d\Et/d\eta$ (more properly $d\Et/d\eta|_{\eta=0}$, measured in an electromagnetic calorimeter and corrected to total hadronic \Et within a reference acceptance of 
$\Delta\eta=1.0, \Delta\phi=2\pi$~\cite{PXppg100}) are presented for $p$$+$$p$, $d$$+$Au and Au$+$Au collisions at \sqsn=200 GeV 
and Au$+$Au collisions at \sqsn=62.4 and 130 GeV. The 
transverse energy \Et is a multiparticle variable defined as the sum
\begin{equation}
   \Et=\sum_i E_i\,\sin\theta_i \qquad d\Et(\eta)/d\eta=\sin\theta(\eta)\, dE(\eta)/d\eta, 
\label{eq:ETdef}
\end{equation}
where $\theta$ is the polar angle, $\eta=-\ln \tan\theta/2$ is the 
pseudorapidity, $E_i$ is by convention taken as the kinetic energy for 
baryons, the kinetic energy + 2 $m_N$ for antibaryons, and the total 
energy for all other particles, and the sum is taken over all particles 
emitted into a fixed solid angle for each event.  

The transverse energy, \Et, was introduced by high energy physicists~\cite{WillisISAproc72,BjorkenPRD8} 
as an improved method to detect and study the jets from hard-scattering compared to high \pt single particle spectra by which hard-scattering was discovered in $p$$+$$p$ collisions and used as a hard-probe in Au$+$Au collisions at RHIC. However, it didn't work as expected: \Et distributions are dominated by soft particles near $\mean{\pt}$~\cite{seebook}.

The significance of systematic measurements of mid-rapidity $d\Et/d\eta$ 
and the closely related charged particle multiplicity distributions, 
$d\Nch/d\eta$, in A$+$B collisions is that 
they provide excellent characterization of the nuclear geometry (hence centrality) of the 
reaction on an event-by-event basis, and are also sensitive to the underlying 
reaction dynamics.  For instance, 
measurements of $d\Nch/d\eta$ in Au$+$Au collisions at the Relativistic 
Heavy Ion Collider (RHIC), as a function of centrality expressed as the 
number of participating nucleons, \Npart, do not depend linearly on \Npart 
but have a nonlinear increase of $\mean{d\Nch/d\eta}$ with increasing 
\Npart. The nonlinearity has been explained by a two component 
model~\cite{WangGyulassyPRL86,KharzeevNardiPLB507} proportional to a 
linear combination of \Ncoll and \Npart, with the implication that the 
\Ncoll term represents a contribution from hard-scattering.  
Alternatively, it has been proposed that $d\Nch/d\eta$ is linearly 
proportional to the number of constituent-quark participants (NQP) 
model~\cite{EreminVoloshinPRC67}, without need to introduce a 
hard-scattering component. For symmetric systems such as Au+Au, the NQP model is identical to another model from the 1980s, the Additive Quark Model (AQM)~\cite{AQMPRD25}, which is actually a model of particle production by color-strings in which only one color-string can be attached to a constituent-quark participant. The models can be distinguished for asymmetric systems such as $d+$Au, since in the AQM the maximum number of color-strings is limited to the number of constituent-quarks in the lighter nucleus, or six for $d+$Au, while the NQP allows all the quark participants in both nuclei to emit particles. 
 The two-component ansatz, the NQP model and the AQM will be tested with the 
present data.
\section{Extreme independent models}
\label{sec:ExtInd}
The models mentioned above are examples of Extreme Independent Models in which the effect of the nuclear geometry of the interaction can be calculated independently of the dynamics of particle production which can be taken directly from experimental measurements. The nuclear geometry is represented by the relative probability, $w_n$ per B$+$A interaction for a given number $n$ of fundamental elements, in the present case, number of collisions (\Ncoll), number of nucleon participants (wounded nucleon model --- WNM~\cite{WNM}), number of constituent-quark participants (\Nqp), number of color-strings (AQM). The dynamics of particle production, the \Nch or \Et distribution of the fundamental element, is taken from the measured $p$$+$$p$ data in the same detector: e.g. the measured \Nch distribution for a $p$$+$$p$ collision represents: 1 collision; 2 participants (WNM); a predictable convolution of constituent-quark participants (NQP), or projectile-quark participants (AQM). Glauber calculations of the nuclear geometry ($w_n$) and the $p$$+$$p$ measurement provide a prediction for the B$+$A measurement in the same detector as the result of particle production by multiple independent fundamental elements.  

\begin{figure}[!b] 
      \centering
      \small
a)\raisebox{-1pc}{\includegraphics[width=0.29\linewidth,height=0.32\linewidth]{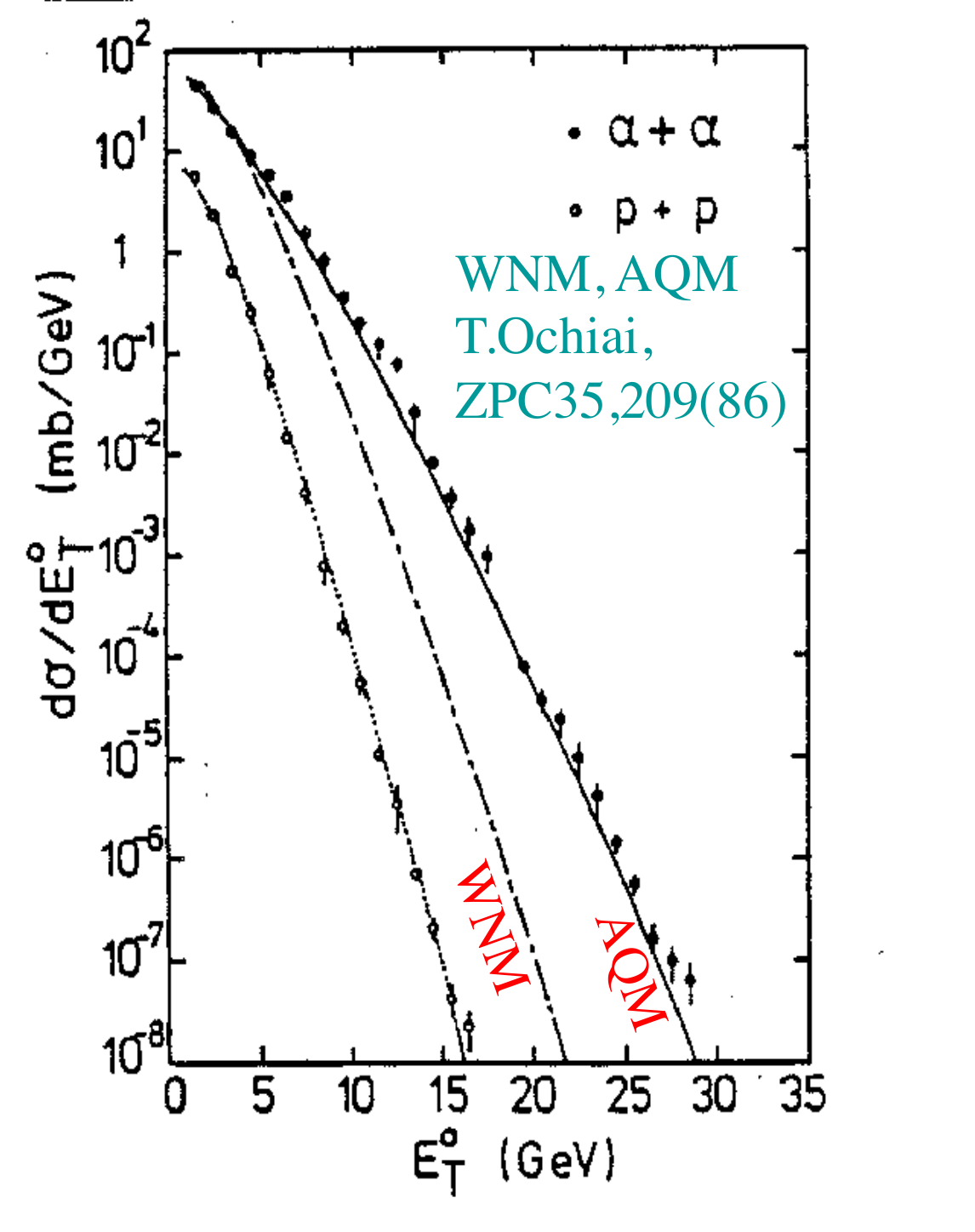}} 
b)\raisebox{-1pc}{\includegraphics[width=0.29\linewidth,height=0.32\linewidth,angle=1]{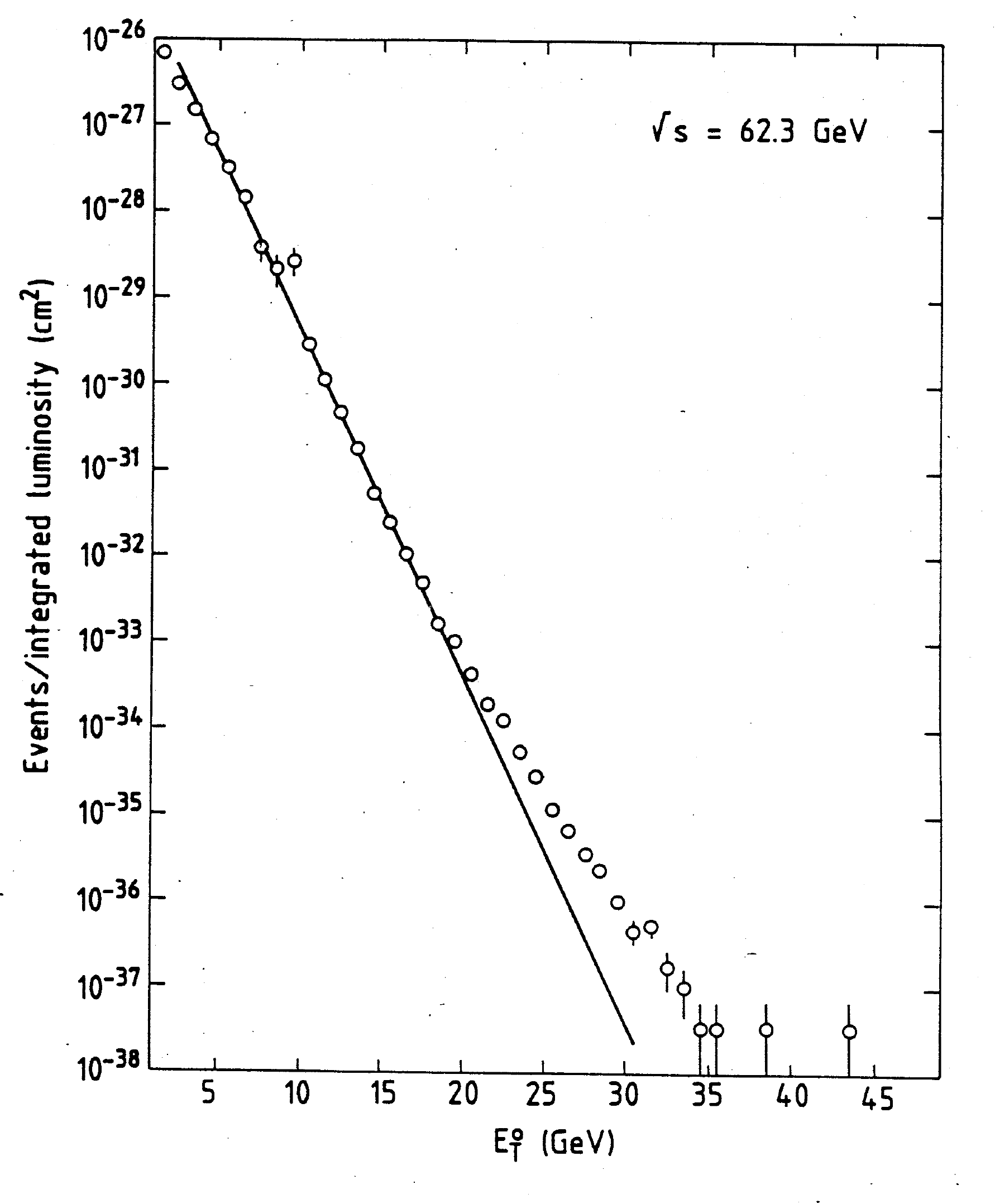}}
c)\raisebox{-0.2pc}{\includegraphics[width=0.29\linewidth,height=0.30\linewidth,angle=-0.0]{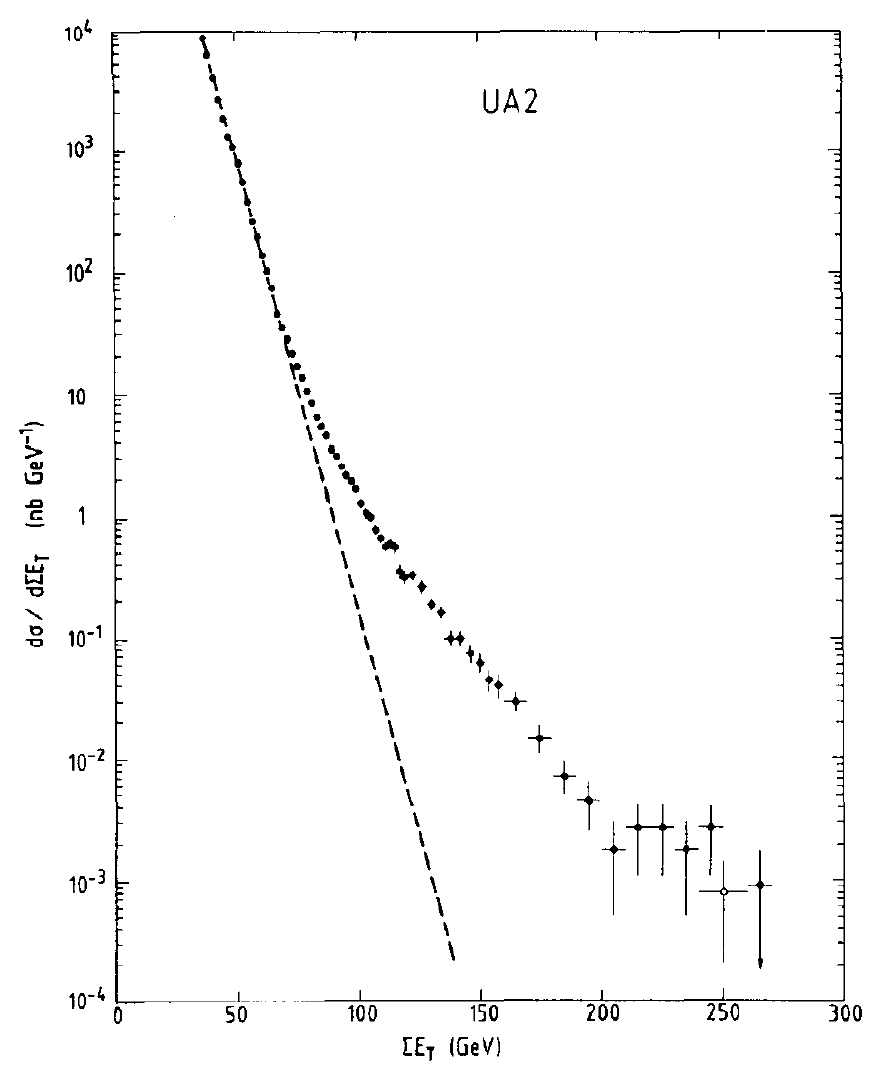}}
\normalsize
     \caption[]{(a) \Et distributions in $p$$+$$p$, $\alpha$$+$$\alpha$~\cite{BCMOR-alfalfa} at \sqsn=31 GeV, with AQM and WNM calculations~\cite{OchiaiZPC35}. (b),(c) \Et distributions with breaks indicating jets: (b) $p$$+$$p$ \sqs=62.3 GeV~\cite{CMORNPB244}; (c) $d\sigma/d\Et$ (nb/GeV) vs. \Et for $\bar{p}$$+$$p$ \sqs=540 GeV~\cite{DiLellaARNPS85}.}
      \label{fig:3plots}
   \end{figure}

I became acquainted with these models in my first Quark Matter talk (QM1984), in which I presented measurements of \Et distributions from $p$$+$$p$ and $\alpha$$+$$\alpha$ interactions at \sqsn=31 GeV at the CERN-ISR (Fig.~\ref{fig:3plots}a~\cite{BCMOR-alfalfa,MJTQM84proc}).  
It was claimed at the meeting that the deviation from the WNM was due to jets~\cite{CallenQM84}, but in both proceedings~\cite{MJTQM84proc,CallenQM84} 
it was demonstrated that ``there is no \ldots\ sign of jets. This indicates that soft processes are still dominant, and that we are still legimately testing the WNM at these high values of \Et.''~\cite{CallenQM84}. As shown in Fig.~\ref{fig:3plots}a, the WNM did not follow the data but the AQM did~\cite{OchiaiZPC35}. Jets do appear in \Et distributions as a break $\lsim10^{-5}$ down in cross section (Figs.~\ref{fig:3plots}b,c).   

\section{\Et and \Nch distributions cut on centrality}
\begin{figure}[!h] 
      \centering
      \small
a)\raisebox{0pc}{\includegraphics[width=0.29\linewidth]{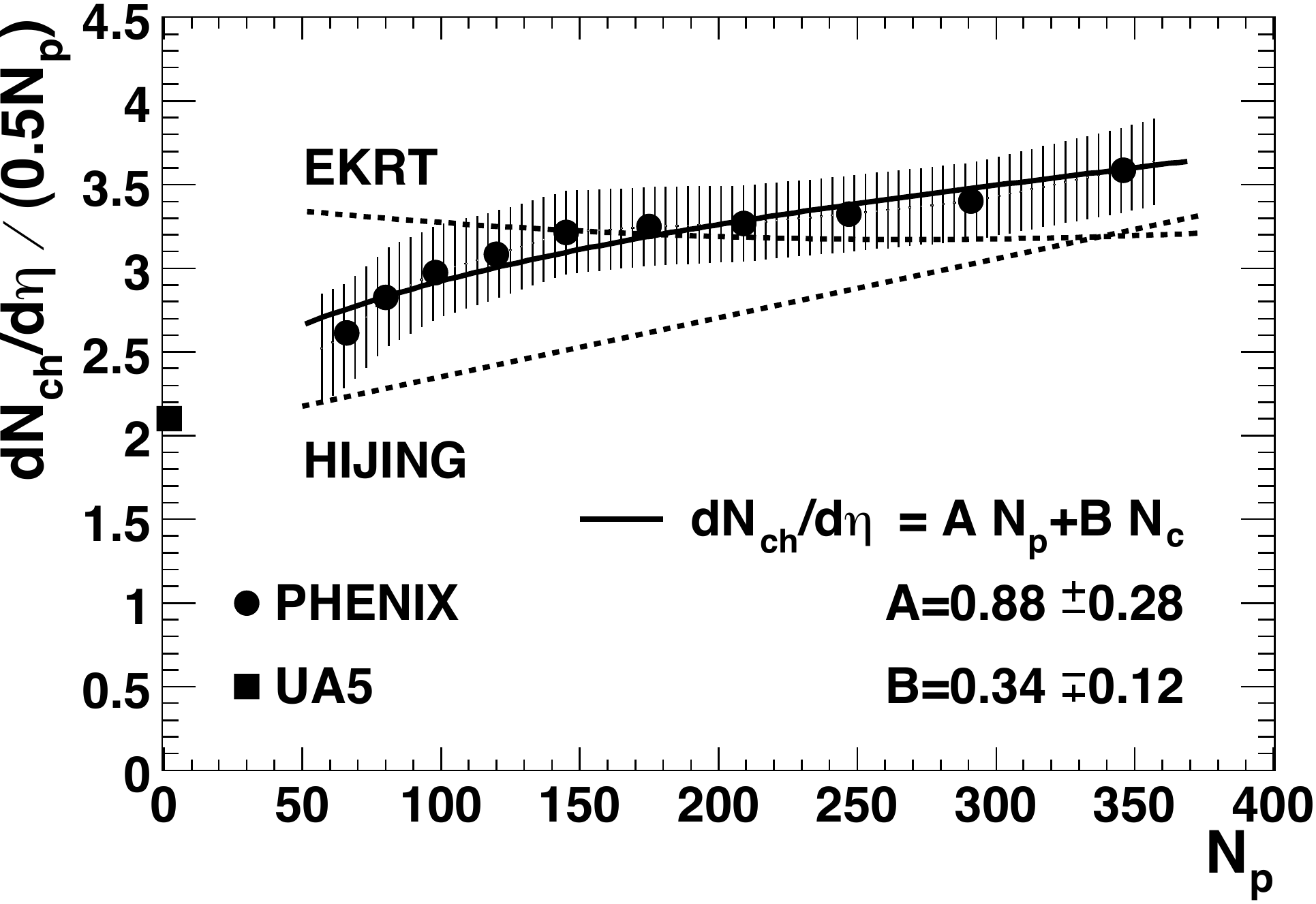}}\hspace*{1pc} 
b)\raisebox{0pc}{\includegraphics[width=0.29\linewidth,height=0.20\linewidth,angle=0]{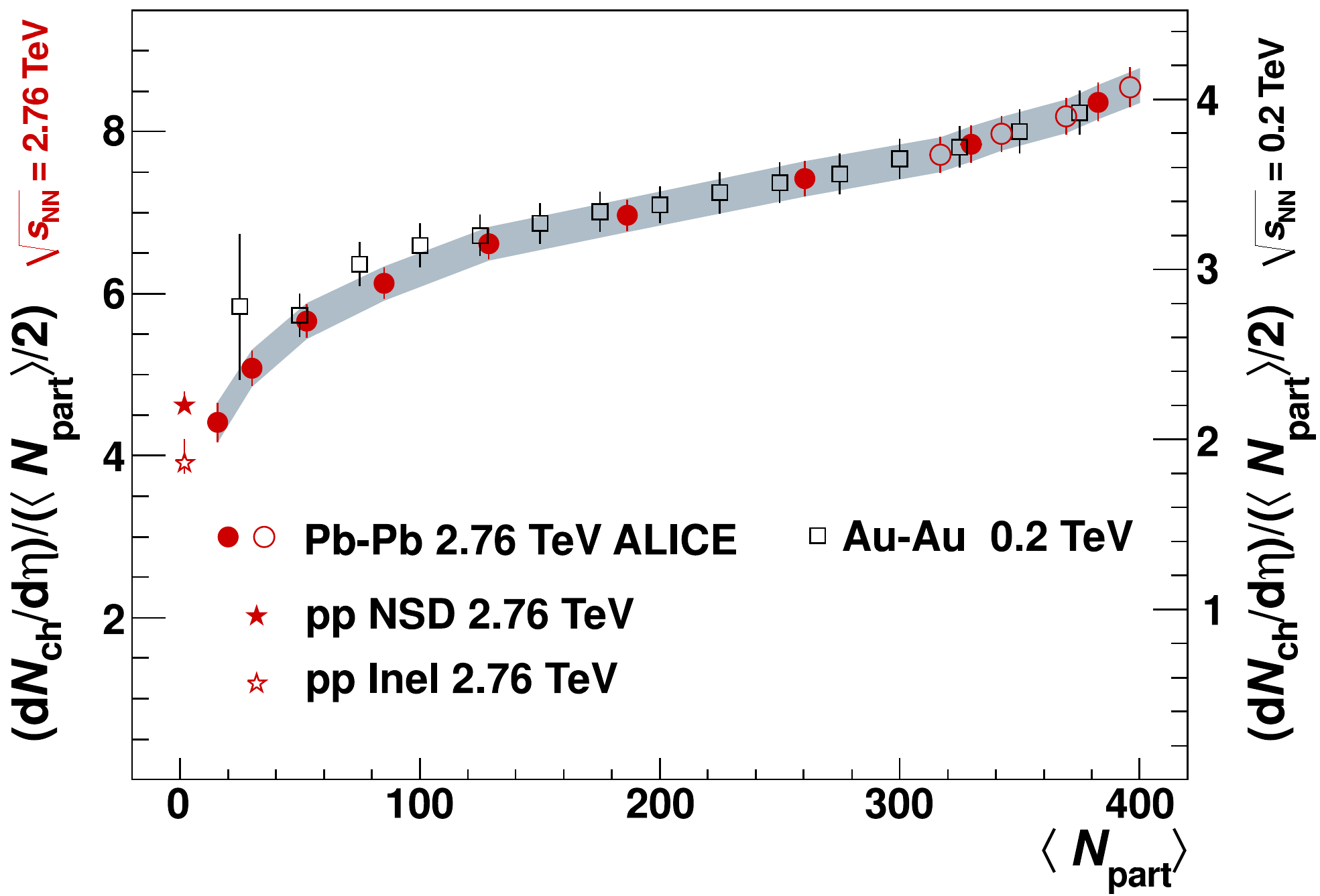}}\hspace*{0.5pc}
c)\raisebox{0pc}{\includegraphics[width=0.29\linewidth,height=0.21\linewidth,angle=-0.0]{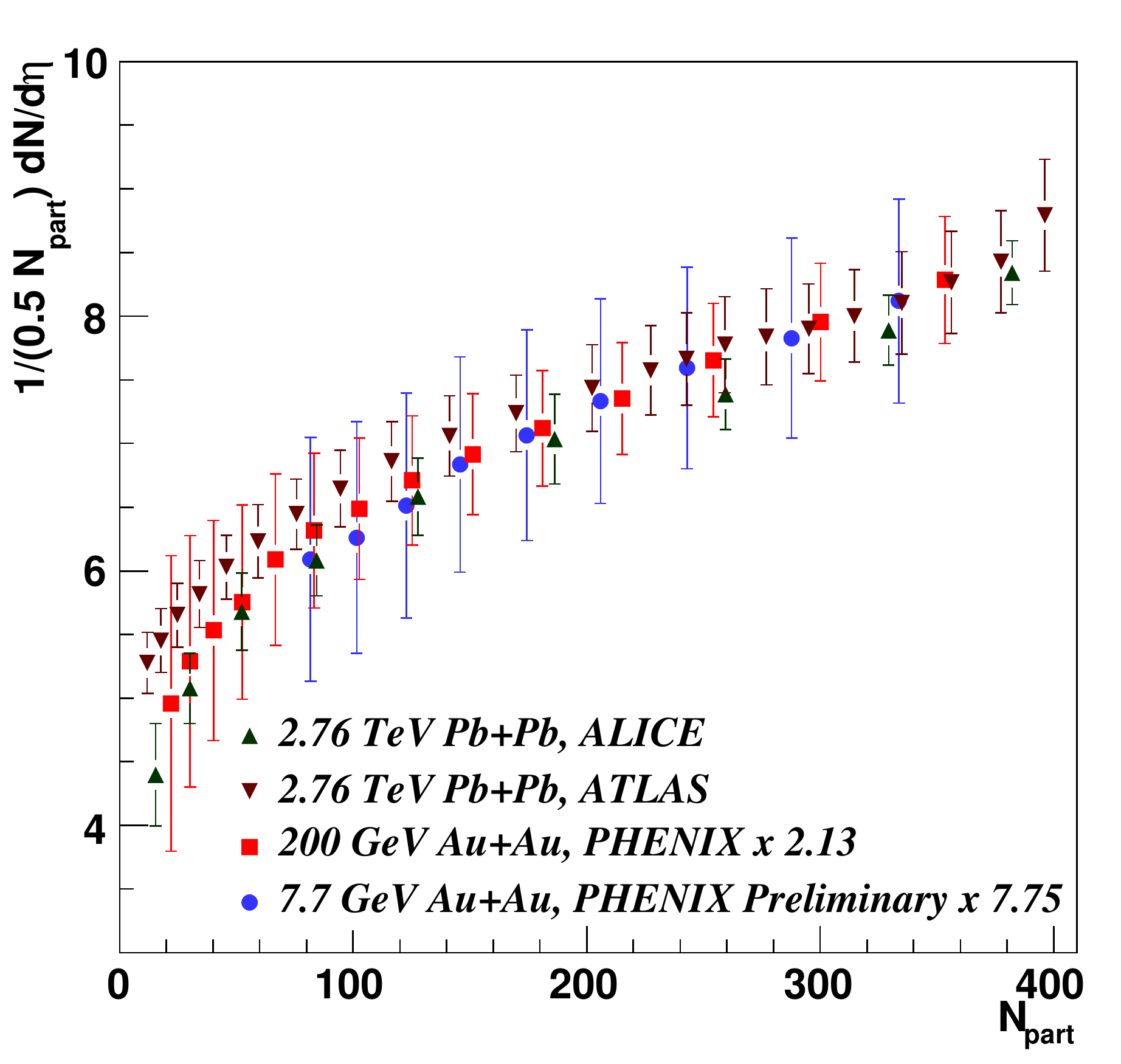}}
\normalsize
     \caption[]{(a) PHENIX, Au$+$Au, \sqsn=130 GeV~\cite{Adcox:2000sp}; (b) ALICE, Pb+Pb, \sqsn=2.76 TeV~\cite{ALICEPRL106}; (c) PHENIX preliminary Au$+$Au, \sqsn=7.7 GeV compared to the data at larger \sqsn~\cite{JTMCPOD8}. }
      \label{fig:3RHICplots}
   \end{figure}

At RHIC, following the style of the CERN SpS rather than the BNL-AGS fixed target heavy ion program, \Et and \Nch distributions were not generally shown. The measurements were presented cut in centrality in the form  $\mean{d\Nch^{\rm AA}/d\eta}/(\mean{\Npart}/2)$ vs. $\mean{\Npart}$ (Fig.~\ref{fig:3RHICplots}), which would be a constant equal to $\mean{d\Nch^{\rm pp}/d\eta}$ if the WNM worked. The measurements clearly deviate from the WNM (Fig.~\ref{fig:3RHICplots}a)~\cite{Adcox:2000sp}; so the PHENIX collaboration, inspired by the    

\noindent preceding article in the journal~\cite{WangGyulassyPRL86}, fit their data to the two-component ansatz:
\begin{equation} 
{d\Nch^{AA}/d\eta}=({d\Nch^{pp}/d\eta})\ [(1-x)\,\mean{\Npart}/2 +x\,\mean{\Ncoll} ] \label{eq:ansatz}
\end{equation}
where the \Ncoll term implied a hard-scattering component for \Et and \Nch, known to be absent in p-p collisions~\footnote{It was noted in Ref.~\cite{Adcox:2000sp} that hard-scattering was not a unique interpretation.  
The shape of the centrality dependences of $d\Et^{AA}/d\eta$ parameterized as $\Npart^\alpha$ were the same for Pb+Pb at \sqsn=17.6 GeV at the CERN SpS and Au$+$Au at \sqsn=130 GeV, with $\alpha=1.1$ and $\alpha=1.16\pm 0.04$, respectively. The LHC data 10 years later~\cite{ALICEPRL106} gave $\alpha=1.19\pm 0.02$ for Pb+Pb at \sqsn=2760 GeV, again the same shape.}
 (recall Fig.~\ref{fig:3plots}). A decade later, the first measurement from Pb+Pb collisions with \sqsn=2.76 TeV at the LHC Fig.~\ref{fig:3RHICplots}b~\cite{ALICEPRL106}, showed exactly the same shape  vs. \Npart as the RHIC Au$+$Au data at \sqsn=200 GeV, although $\mean{\Ncoll}$ is a factor of 1.6 larger and the hard-scattering cross section is more than a factor of 20 larger. This strongly argued against a hard-scattering component and for a nuclear geometrical effect, which was reinforced by a PHENIX preliminary measurement in Au$+$Au at \sqsn=7.7 GeV (Fig.~\ref{fig:3RHICplots}c)~\cite{JTMCPOD8} which also showed the same shape for the evolution of $\mean{d\Nch^{\rm AA}/d\eta}/(\mean{\Npart}/2)$ with \Npart as the \sqsn=200 and 2760 GeV measurements. 
It had previously been proposed that the number of constituent-quark participants provided the nuclear geometry that could explain the RHIC Au$+$Au data without the need to introduce a hard-scattering component~\cite{EreminVoloshinPRC67}.  However an asymmetric system is necessary in order to distinguish the NQP model from the AQM so the two models were applied to the RHIC $d$$+$Au data. \vspace*{-1pc}
\section{The number of constituent-quark participant model (NQP)}
The massive constituent-quarks~\cite{MGMquark}, which form mesons and nucleons (e.g. a proton=$uud$), are relevant for static properties and soft physics with $p_T<2$ GeV/c. They are complex objects or quasiparticles~\cite{ShuryakNPB203} made of the massless partons (valence quarks, gluons and sea quarks) of DIS~\cite{DIS2} such that the valence quarks acquire masses $\approx 1/3$ the nucleon mass with radii $\approx 0.3$ fm when bound in the nucleon. With  smaller resolution one can see inside the bag to resolve the massless partons which can scatter at large angles according to QCD. At RHIC, hard-scattering is distinguishable from soft (exponential) particle production only for $p_T\geq$ 2 GeV/c at mid-rapidity, where $Q^2=2p_T^2=8$ (GeV/c)$^2$ which corresponds to a distance scale (resolution) $<0.07$ fm.

\begin{figure}[!h] 
      \centering
\raisebox{-0.3pc}{\includegraphics[width=0.29\linewidth,height=0.21\linewidth]{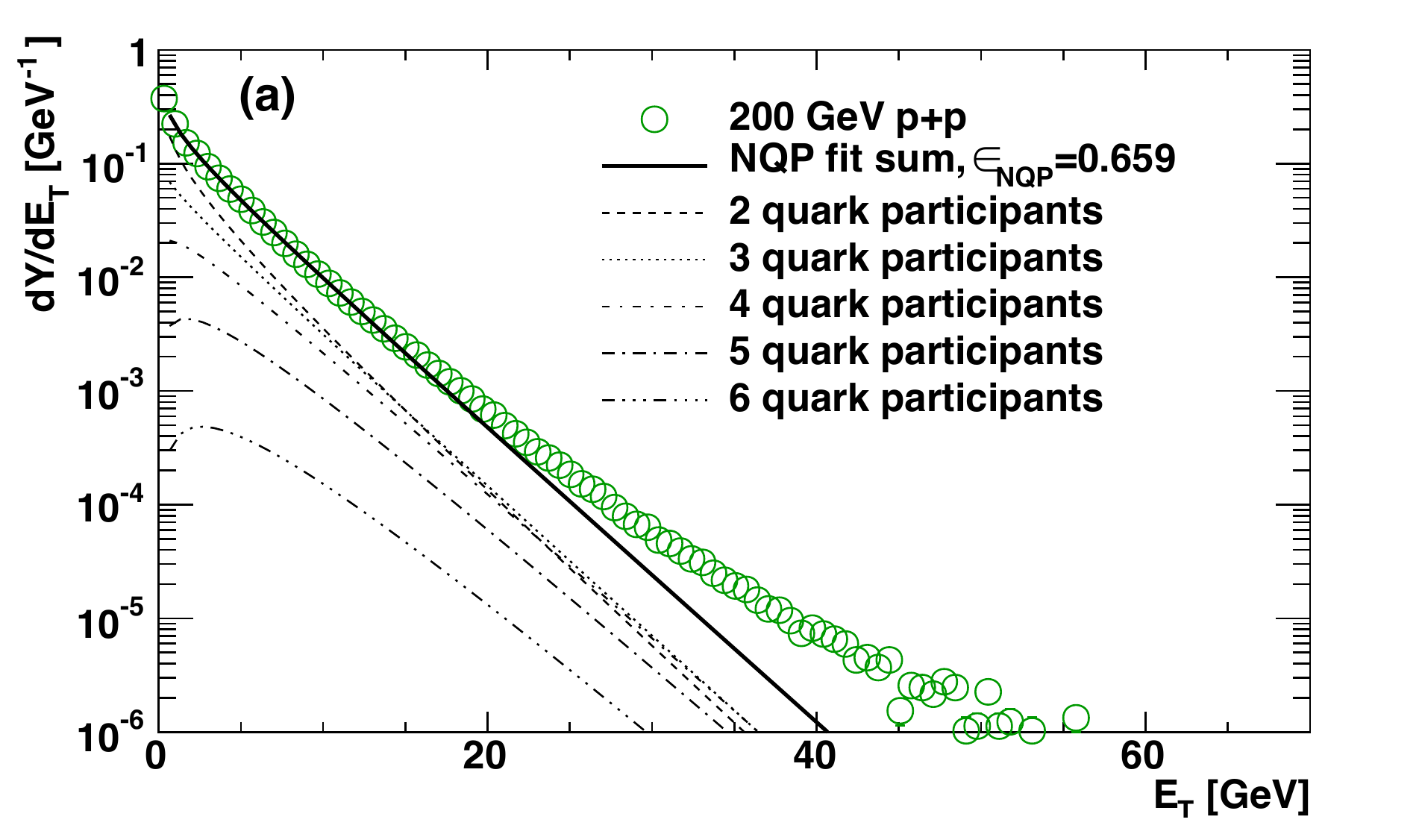}}\hspace*{1pc} 
\raisebox{0pc}{\includegraphics[width=0.29\linewidth,height=0.21\linewidth,angle=0]{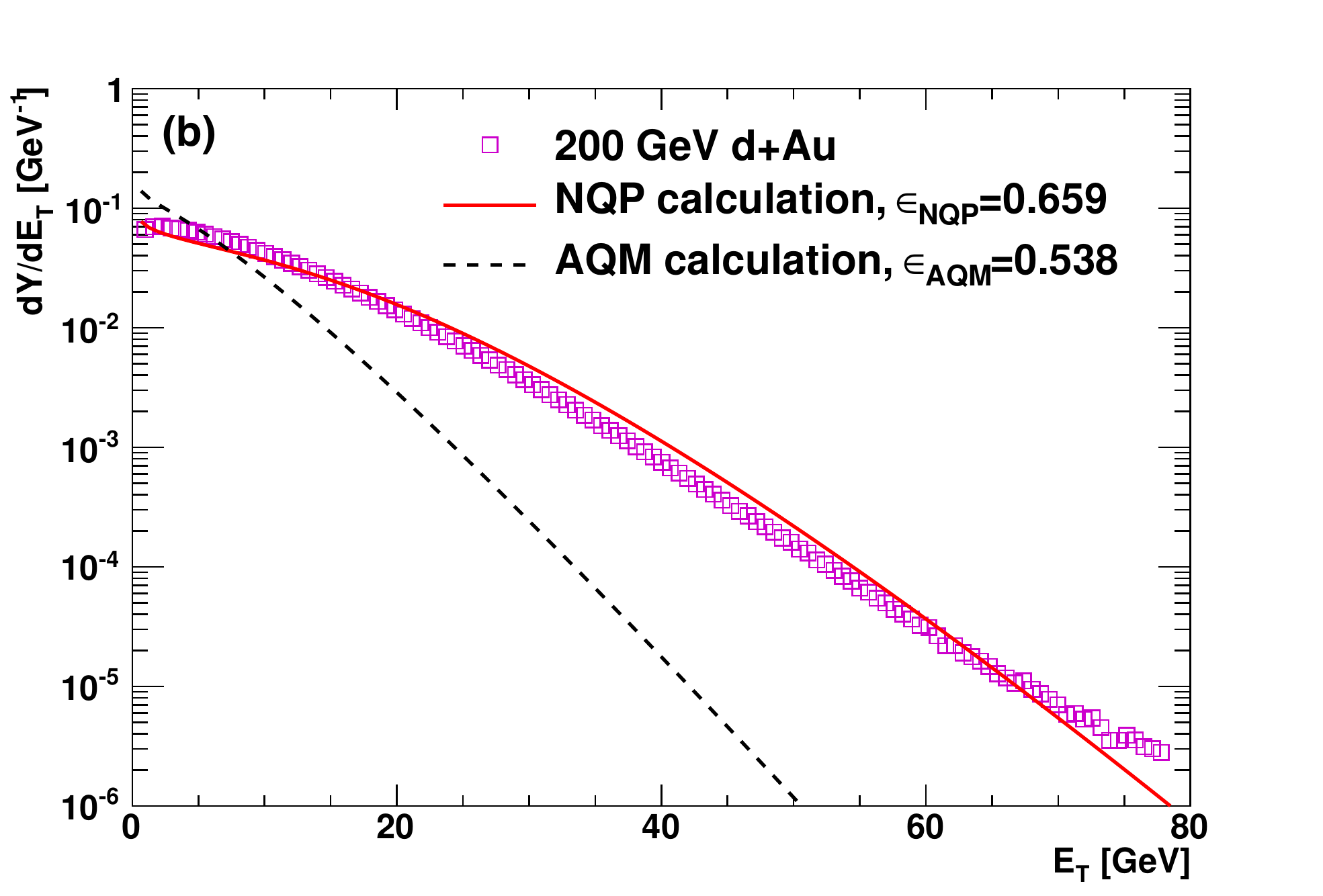}}\hspace*{0.5pc}
\raisebox{0pc}{\includegraphics[width=0.29\linewidth,height=0.21\linewidth,angle=-0.0]{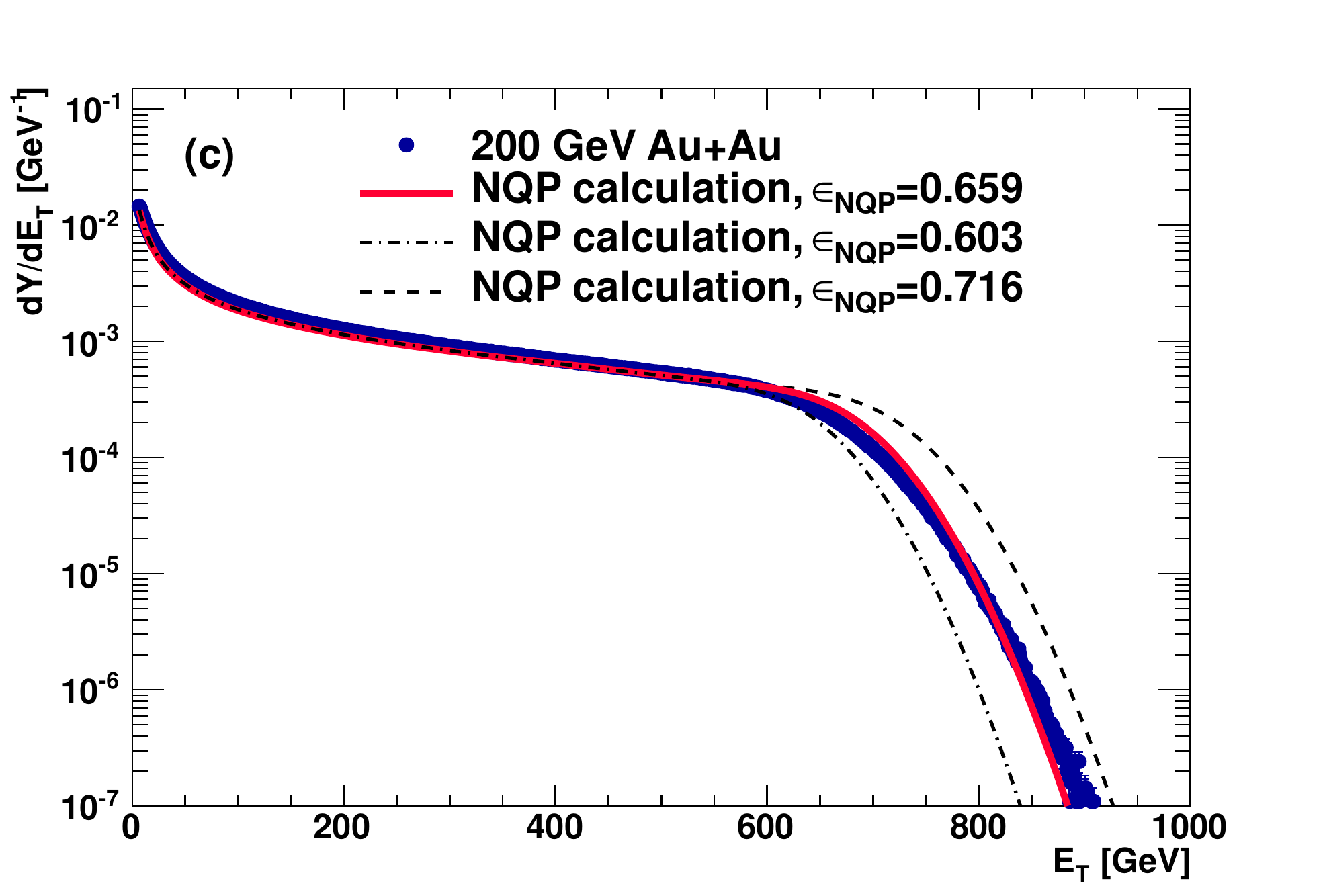}}
\normalsize \vspace*{-0.25pc}
     \caption[]{PHENIX NQP calculations~\cite{PXppg100} for (a) $p$$+$$p$, (b) $d$$+$Au (also AQM), (c) Au$+$Au \Et distributions at \sqsn=200 GeV.}
      \label{fig:NQPcalc}\vspace*{-0.5pc}
   \end{figure}

A standard Monte Carlo Glauber calculation is used to assemble the initial positions of all the nucleons. Then three quarks are distributed around the center of each nucleon according to the proton charge distribution \mbox{$\rho(\vec{r})\propto e^{-{\sqrt{12}\,r/r_{\rm rms}}}$}, where $r_{\rm rms}=0.81$ fm is the rms charge radius of the proton~\cite{HofstadterRMP}. The $q$--$q$ inelastic scattering cross section is adjusted to 9.36 mb  at \sqs=200 GeV to give the correct $p$$+$$p$ inelastic cross section (42 mb)  and then used in the B$+$A calculations. Fig.~\ref{fig:NQPcalc}a shows the deconvolution of the $p$$+$$p$ \Et distribution to the sum of 2--6 constituent-quark participants from which the \Et distribution of a constituent-quark is determined and applied to $d$$+$Au (Fig.~\ref{fig:NQPcalc}b) and Au$+$Au (Fig.~\ref{fig:NQPcalc}c) reactions in the same detector. The NQP calculations closely follow the measured $d$$+$Au and Au$+$Au \Et distributions in shape and magnitude over a range of more than 1000 in cross section. A complete calculation was also done for the AQM which fails to describe the $d$$+$Au data (Fig.~\ref{fig:NQPcalc}b). The conclusion is that it is the number of constituent-quark participants that determine the \Nch and \Et distributions and that the AQM calculation which describes the $\alpha$--$\alpha$ data at \sqsn=31 GeV (Fig.~\ref{fig:3plots}a) is equivalent to the NQP in the symmetric system. 

\begin{figure}[!h] 
      \centering
            \small
a)\raisebox{0.0pc}{\includegraphics[width=0.32\linewidth,height=0.22\linewidth]{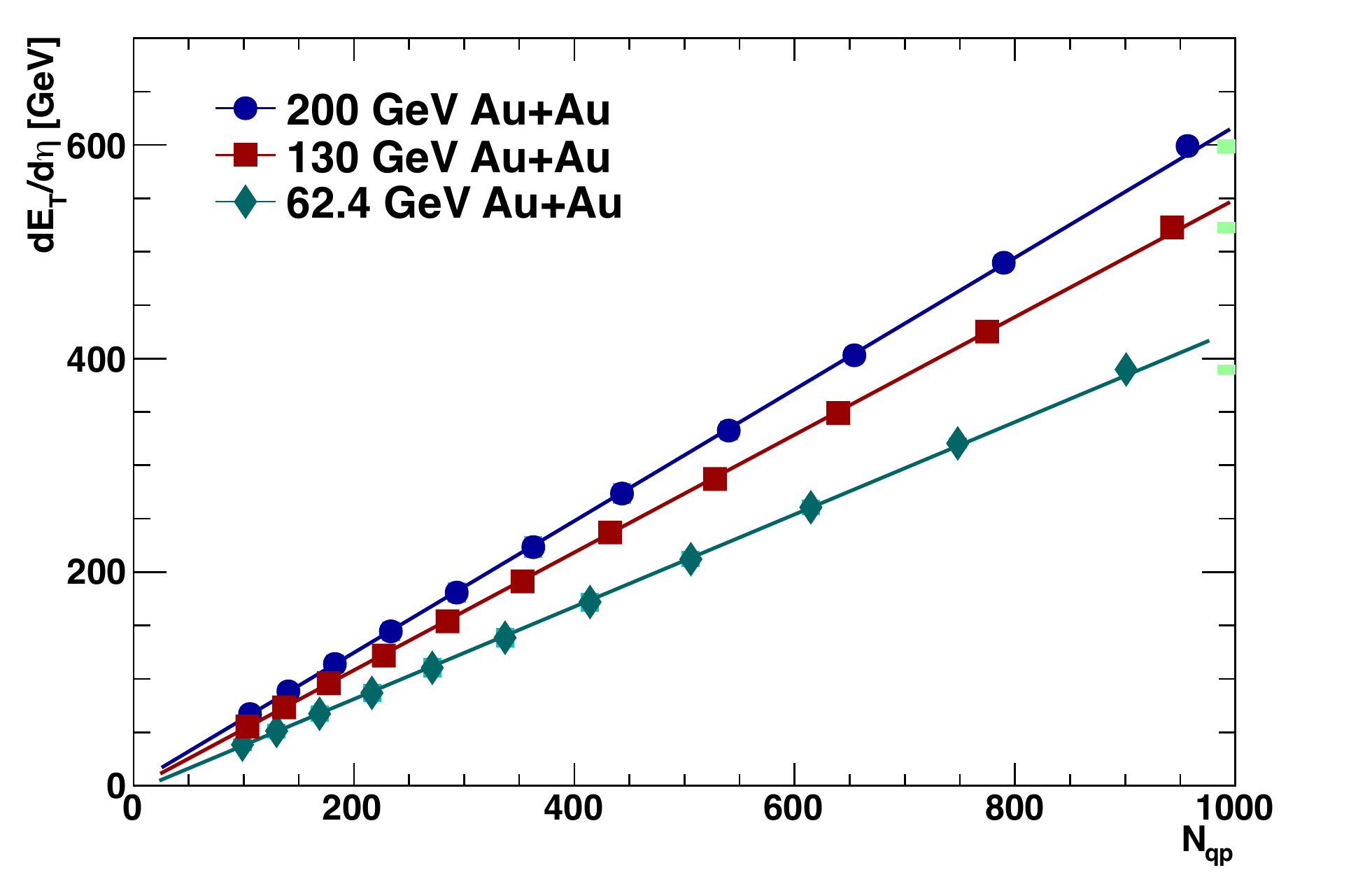}}
b)\raisebox{0pc}{\includegraphics[width=0.32\linewidth,height=0.22\linewidth,angle=0]{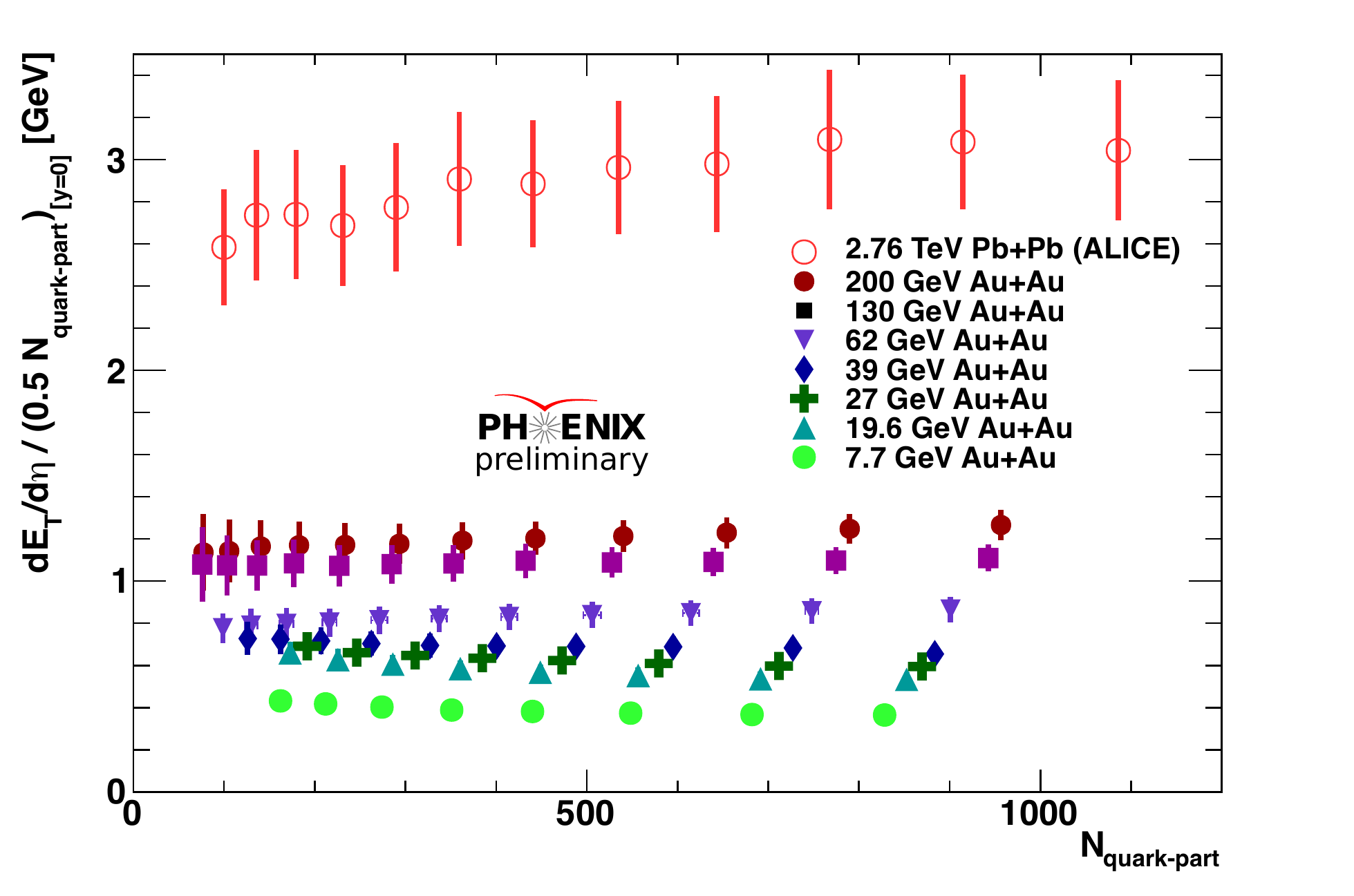}}\hspace*{-0.2pc}
c)\raisebox{-0.5pc}{\includegraphics[width=0.32\linewidth,height=0.23\linewidth,angle=-0.0]{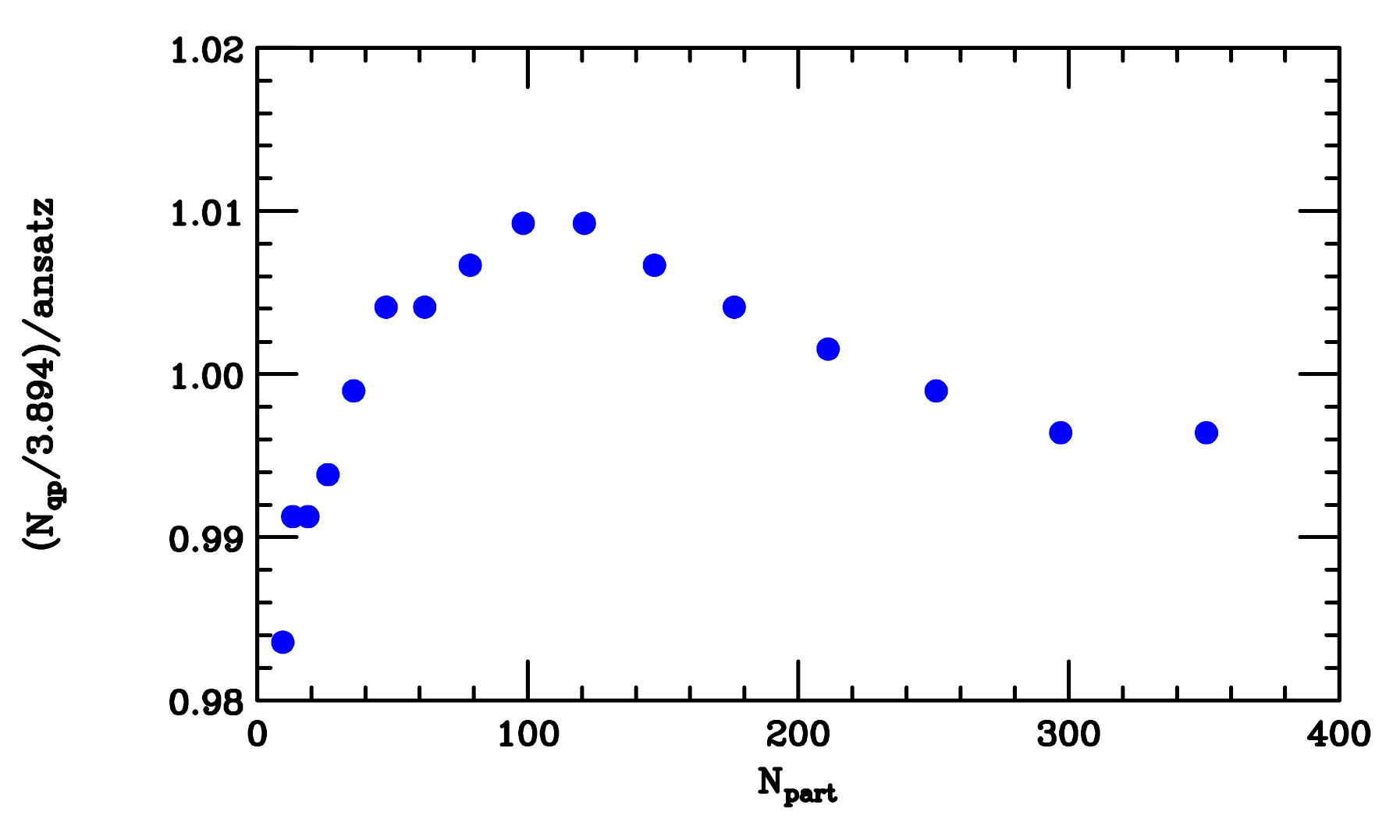}}
\normalsize\vspace*{-1.25pc}
     \caption[]{PHENIX~\cite{PXppg100}: (a) $d\Et/d\eta$ vs. \Nqp;  (b) $d\Et/d\eta/(\Nqp/2)$ vs. \Nqp;  (c) (\Nqp/ansatz)/$\mean{\Nqp/{\rm ansatz}}$ vs. \Npart. }
      \label{fig:NQPcent}
   \end{figure}

\noindent The NQP model was also applied to the centrality cut PHENIX data with the result that $d\Et/d\eta$ is strictly proportional to \Nqp (Fig~\ref{fig:NQPcent}a) so that $d\Et/d\eta/(\Nqp/2)$ vs. \Nqp is a constant for $\sqsn\, \gsim\, 27$ GeV (Fig~\ref{fig:NQPcent}b). As a final touch, the ratio of the two-component ansatz with $x=0.08$ to the \Nqp as a function of centrality was found to be constant to better than 1\% at \sqsn=200 GeV (Fig~\ref{fig:NQPcent}c), which indicates that the ansatz works because the particular linear combination of \Npart and \Ncoll turns out to be an empirical proxy for \Nqp and not because the \Ncoll term implies hard-scattering.
\vspace*{-1.0pc}








\end{document}